\documentclass[prd,twocolumn,superscriptaddress,showpacs,nofootinbib,preprintnumbers]{revtex4}

\usepackage{amsmath}
\usepackage{graphicx,bm}% Include figure files
\usepackage{dcolumn}    % Align table columns on decimal point

\begin{document}
\title{Generating the primordial curvature perturbations in preheating}
\author{Teruaki Suyama}
\affiliation{Department of Physics, Kyoto University, Kyoto 606-8502, Japan}

\author{Shuichiro Yokoyama}
\affiliation{Department of Physics, Kyoto University, Kyoto 606-8502, Japan}

\preprint{KUNS-2030}
\date{today}

\begin{abstract}
We show that the primordial curvature perturbations may originate not from the 
quantum fluctuations of inflaton during inflation but from isocurvature
perturbations which are amplified during preheating after inflation.
We consider a simple preheating model,
whose potential is given by $V(\phi, \chi)=\frac{1}{4}\lambda \phi^4+\frac{1}{2}g^2 \phi^2 \chi^2$
with $g^2/\lambda=2$,
as a possible realization of generating curvature perturbations during preheating.
We make use of the $\delta N$ formalism which requires only knowledge of the 
homogeneous background solutions in order to evaluate the evolution of curvature 
perturbations on super-horizon scales.
We solve the background equations numerically and find that the amplitude and the 
spectral index of curvature perturbations originating from preheating 
can be tuned to the observed values if the isocurvature
perturbations at the end of inflation are not suppressed on super-horizon scales.
We also point out that the tensor to scalar ratio in $\frac{1}{4}\lambda \phi^4$ 
inflation model can be significantly lowered,
hence letting the $\frac{1}{4}\lambda \phi^4$ model, 
which is ruled out by the Wilkinson Microwave Anisotropy 
Probe (WMAP) data combined with SDSS data, 
get back into the observationally allowed region.
\end{abstract}


\maketitle

\section{Introduction}
The primordial curvature perturbations which are coherent over the horizon scales
at recombination epoch are seeds for large scale structure.
A simple scenario to explain the origin of curvature perturbations is that 
vacuum fluctuations of inflaton driving inflation convert into classical 
perturbations at around the time of horizon crossing, 
which are almost adiabatic, 
almost scale-invariant and almost Gaussian.
These perturbations are constant until the scales of interest re-enter the horizon.

Recently alternative mechanisms were proposed which generate the primordial
curvature perturbations after inflation,
not from the inflaton fluctuations.  
In curvaton scenario \cite{Linde:1996gt, Moroi:2001ct, Lyth:2001nq},
quantum fluctuations of light scalar field during inflation which are not necessarily  
related to the inflaton, convert into curvature perturbations after inflation ends.
In inhomogeneous reheating scenario \cite{Dvali:2003em, Kofman:2003nx, Matarrese:2003tk},
spatial fluctuations of the decay rate of the inflaton to radiation generate
curvature perturbations.
Both of these mechanisms are based on the formula \cite{Lyth}
\begin{equation}
{\dot {\cal R}_c}=\frac{H}{\rho+P}\delta P_{\rm nad},  \label{int1}
\end{equation}
which is valid on super-horizon scales.
Here ~${\dot {}}$ ~is a 
derivative with respect to the cosmological time,
$H \equiv \dot{a}/a$ represents the Hubble parameter, $\rho$ is the energy density, $P$ is the pressure, 
${\cal R}_c$ is the curvature perturbation on comoving hypersurface and
$\delta P_{\rm nad} \equiv \delta P-\frac{{\dot P}}{ {\dot \rho}} \delta \rho$
is called the non-adiabatic part of the pressure perturbation or isocurvature perturbation.
If $\delta P_{\rm nad}$ is non-zero after inflation,
which is the case of the curvaton and the inhomogeneous reheating scenarios,
${\cal R}_{c}$ on super-horizon scales is generated after inflation.

As another candidate for the mechanism of generating curvature perturbations
after inflation,
preheating model has also been considered~\cite{Bassett:1999mt, Bassett:1998wg,
Zibin:2000uw, Tanaka:2003ck, Kolb:2004jm}.
Preheating \cite{Traschen:1990sw, Kofman:1994rk, Yoshimura:1995gc, Fujisaki:1995dy} is a process 
in which the coherent oscillation of the inflaton excites the fluctuations of 
fields coupled to the inflaton by parametric resonance. 
This process was investigated in detail \cite{Yoshimura:1995gc, Fujisaki:1995dy, Son:1996uv, Boyanovsky:1996sq, Prokopec:1996rr, Baacke:1996kj, Khlebnikov:1996mc, Khlebnikov:1996wr, Khlebnikov:1996zt, Boyanovsky:1997cr, Kofman:1997yn, Greene:1997fu, Baacke:1997rs, Greene:1997ge, Greene:1998nh, Suyama:2004mz}
without metric perturbations,
which is valid in dealing processes occurring well within the horizon scale.
It was then recognized that particle production in preheating occurs much more
efficiently than in the old reheating scenario where inflaton decays perturbatively 
to other particles.

Naively, 
fluctuations of such fields coupled to the inflaton seem to be excellent source 
of isocurvature perturbations.
This means there is a possibility of generating curvature perturbations on 
super-horizon scales during preheating epoch.
However, whether preheating can affect the evolution of curvature perturbations 
on super-horizon scales or not \cite{Bassett:1999mt, Bassett:1998wg} is not 
completely understood yet.
A necessary condition that curvature perturbations on super-horizon scales
are affected by parametric resonance is that isocurvature perturbations are
not suppressed on super-horizon scales \cite{Gordon:2000hv}.
Though this condition is not satisfied for large class of two-field preheating models,
there remains some models which satisfy this condition.
A simple example is a massless self-coupled inflaton $\phi$ coupled to another scalar
field $\chi$,
i.e. with potential \cite{Bassett:1999cg}
\begin{equation}
V(\phi, \chi)=\frac{\lambda}{4}\phi^4+\frac{g^2}{2}\phi^2 \chi^2, \label{int2}
\end{equation}
with $g^2/\lambda = 1\sim 3$.

For the model given in Eq.~(\ref{int2}) with $g^2/\lambda=2$,
where the $k=0$ mode has the largest growth rate($k$ is the wavenumber),
it was suggested in Ref.~\cite{Zibin:2000uw} that large-scale curvature perturbations 
are much amplified and exceed the current observational upper bound by using the
mean field approximation.

On the other hand,
in Ref.~\cite{Tanaka:2003ck},
by making use of the $\delta N$ formalism \cite{Sasaki:1995aw, Sasaki:1998ug, Wands:2000dp},
it was suggested 
that the amplification of curvature perturbations during preheating is less 
efficient than is expected in \cite{Bassett:1999cg, Zibin:2000uw}.

In the $\delta N$ formalism,
we can evaluate the evolution of curvature perturbations only by the knowledge
of the homogeneous background solutions and as was emphasized in \cite{Tanaka:2003ck},
unphysical acausal energy transfer does not occur in $\delta N$ formalism while
this energy transfer is not inhibited in the mean field approximation
used in \cite{Zibin:2000uw}.

However in Ref.~\cite{Tanaka:2003ck},
whether the generation of curvature perturbations on super-horizon scales takes 
place during preheating or not remains an open question because the number of solving
background equations during preheating varying the initial value of the isocurvature
perturbations at the end of inflation is not enough.

In this paper,
we numerically solve background equations for the model given by Eq.~(\ref{int2}) 
during preheating and show that scale invariant curvature perturbations with amplitude 
observed in the temperature anisotropy of Cosmic Microwave Background(CMB) 
can be generated during preheating.
\footnote{Generating the primordial curvature perturbations in preheating was also 
discussed in \cite{Kolb:2004jm}.
They considered instant preheating model \cite{Felder:1998vq} in which case the resulting
curvature perturbations are proportional to the initial isocurvature perturbations.
As we will show in Sec. IV,
the resulting curvature perturbations for the case we consider are very sensitive 
to the tiny differences of the initial isocurvature perturbations.}

Before we end this section,
we define the power spectrum of any field $f({\vec x})$,
which we denote as ${\cal P}_f$,
by
\begin{equation}
\langle f_{\vec k_1} f_{\vec k_2} \rangle =\frac{2\pi^2}{k^3_1} {\cal P}_f (k_1) \delta ( {\vec k_1}+{\vec k_2} ). \label{int3}
\end{equation}
Here $\langle \cdots \rangle$ is the ensemble average and $\vec{k}$ is the wave number vector.

The paper is organized as follows.
In Sec. II we briefly review the $\delta N$ formalism which we use in a later section
in order to evaluate curvature perturbations generated during preheating.
In Sec. III we give the basics of preheating and give homogeneous background equations
which we numerically solve.
In Sec. IV we solve the background equations numerically and discuss how we can obtain
the power spectrum of curvature perturbations from the solutions of the background
equations.
Then we give the resulting power spectrum after preheating.
Sec. V is a summary.

\section{$\delta N$ formalism}
\label{sepsect}

In this section, 
we briefly review the $\delta N$ formalism~\cite{Sasaki:1995aw, Sasaki:1998ug, Wands:2000dp}.

\begin{figure}[t]
\begin{center}
  \includegraphics[width=8.cm,clip]{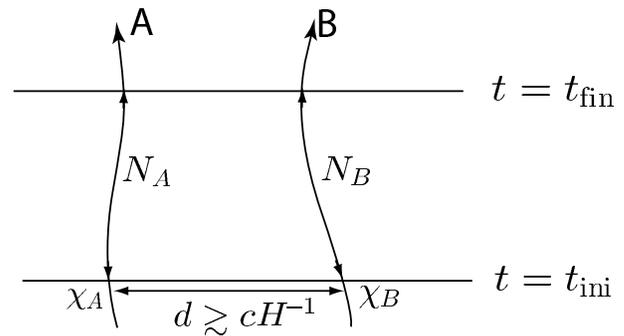}
\caption{ The schematic picture of the $\delta N$ formalism
}
\label{sepfig}
\end{center} 
\end{figure}

Let us consider two world lines $A$ and $B$ which represent the evolution of 
two separate homogeneous universes
separated by the distance $d$,
and take $t_{\rm ini}$ to be a certain time soon after the relevant scale 
$d$, 
becomes larger than the horizon scale
$cH^{-1}$,
and $t_{\rm fin}$ to be a certain time after the complete reheating.

As is well known, 
at $t>t_{\rm fin}$ the curvature perturbation on comoving hypersurface, 
${\cal R}_{c}$ becomes constant in time on super-horizon scales. 
We define the background $e$-folding number as 
\begin{eqnarray}
N \equiv \int^{t_{\rm fin}}_{t_{\rm ini}} Hdt~.
\end{eqnarray}
According to \cite{Sasaki:1995aw, Sasaki:1998ug, Wands:2000dp}, 
 taking the initial $(t=t_{\rm ini})$ hypersurface to be flat and the final 
$(t=t_{\rm fin})$ hypersurface to be comoving, 
there is a simple relation between ${\cal R}_c$ and the perturbation of the 
background $e$-folding number as
\begin{eqnarray}
\label{dpsi2}
{\cal R}_{cd}(t_{\rm fin}) &=& - \left[ N_A(t_{\rm fin},t_{\rm ini})- N_B(t_{\rm fin},t_{\rm ini})
\right]
\, ,
\end{eqnarray}
where ${\cal R}_{cd}$ represents the comoving curvature perturbation on super-horizon scale, $d$,
and $N_{A(B)}$ represents the $e$-folding number measured from $t_{\rm ini}$ to $t_{\rm fin}$ along the 
world line $A(B)$. 
Thus,
in order to evaluate the final curvature perturbation on super-horizon scales, 
it is only necessary to calculate the difference in the $e$-folding number measured 
from the initial to final hypersurfaces between separate FRW universes.
It should be also noted that we only assume that the metric perturbations on 
super-horizon scales are linear perturbations,
while we do not assume the linearity of the matter fields in the FRW universes here.
Hence we can also use this formula for the case in which the dynamics of matter 
fields is non-linear, 
such as in preheating.

In our calculation, 
we obtain the $e$-folding numbers of separate homogeneous universes from 
$t_{\rm ini}$ to $t_{\rm fin}$,
given the different values of $\chi$ on initial hypersurface at $t=t_{\rm ini}$.
We take  the $t=t_{\rm fin}$ 
hypersurface to be constant Hubble one, 
that is, 
$H_A=H_B$, 
rather than comoving hypersurface because it is much easier to handle in the numerical
calculations.
It is well known that the constant Hubble hypersurface is identical to the comoving 
hypersurface after complete reheating, 
that is,
during the universe is characterized by a single parameter, 
e.g., 
the energy density of the radiation. 

\section{Preheating Model}
\subsection{Model}
We consider a two-field preheating model whose effective potential is given by
Eq.~(\ref{int2}).
In the linear stage of preheating where fluctuations of both $\phi$ and
$\chi$ have negligible effects on the background dynamics,
the homogeneous part of $\phi$ oscillates under the potential $\frac{\lambda}{4}\phi^4$.
The explicit form of the oscillation is $a^{-1} \phi_o {\rm cn} (x; \frac{1}{\sqrt{2}})$,
where $a$ is the scale factor,
$\phi_o$ is the initial amplitude of the oscillation of $\phi$, 
$x \equiv \sqrt{\lambda}\phi_o \eta$ ($\eta$ is a conformal time) and
${\rm cn}$ is the Jacobi cosine function.
In this regime,
the evolution equation for a given mode $k$ of the $\chi$ field is given by~\cite{Greene:1997fu} 
\begin{equation}
X_k''+\left( \kappa^2+\frac{g^2}{\lambda} {\rm cn}^2 \left( x;\frac{1}{\sqrt{2}} \right) \right) X_k=0, \label{model2}
\end{equation}
where $X_k \equiv a \chi_k, \kappa^2 \equiv k^2/(\lambda \phi^2_o)$
and $'$ denotes the derivative with respect to $x$.
It is known that for $n(2n-1) <g^2 /\lambda< n(2n+1)$
(n is a positive integer), 
Eq.~(\ref{model2}) has a solution which grows exponentially at $\kappa=0$, i.e., in the long wavelength limit.
Interestingly,
for $g^2/\lambda=2$,
the longest wavelength mode has the largest growth rate and the power spectrum 
of $\chi$-field at the end of inflation is scale-invariant \cite{Zibin:2000uw}
if the interaction between $\phi$ and $\chi$ during inflation is also given
by Eq.~(\ref{int2}).

If we allow the non-minimal negative coupling $\frac{1}{2}\xi R \chi^2$ 
as an additional interaction to the effective potential Eq.~(\ref{int2}) \cite{Tsujikawa:2002nf},
the effective mass squared of $\chi$-field is 
\begin{equation}
m^2_{\chi} \approx g^2 \phi^2+8\pi \lambda \xi \phi^2 {\left( \frac{\phi}{M_p} \right)}^2. \label{model3}
\end{equation}
Since the second term decreases faster than the first one as $\phi$ decreases,
it is possible that the second term dominates during inflation but becomes 
negligible during preheating.
In this case,
the power spectrum of $\chi$ at the end of inflation can be red rather than 
scale-invariant.
In the following discussions, 
we do not fix the initial power spectrum of $\chi$-field and study the resulting 
power spectrum of the curvature perturbation after preheating for various 
cases of ${\cal P}_{\chi}$.

At the linear stage of preheating,
$\chi$-field which corresponds to the isocurvature perturbation \cite{Gordon:2000hv}
grows exponentially by parametric resonance ($\phi$-field fluctuations also 
grow but its growth rate is much smaller than that of $\chi$-field.\cite{Greene:1997fu})
Eventually non-linear interactions of fields dominate the dynamics.
Thus, 
in order to evaluate curvature perturbations, 
we use the $\delta N$ formalism instead of solving the perturbation equations. 

\subsection{Background equations}
In the two-field preheating model in which the effective potential is given by 
Eq.~(\ref{int2}),
the homogeneous background equations are  
\begin{eqnarray}
&&\ddot{\phi}+(3H+\Gamma)\dot{\phi}+\lambda \phi^3 +g^2 \phi \chi^2=0, \label{model4} \\
&&\ddot{\chi}+(3H+\Gamma)\dot{\chi}+g^2 \phi^2 \chi=0, \label{model5} \\
&&\dot{\rho_R}+4H\rho_R=\Gamma (\dot{\phi}^2+\dot{\chi}^2), \label{model6} \\
&&H^2=\frac{8\pi}{3 M^2_p} \left( \rho_S +\rho_R \right), \label{model7} \\
&&\rho_S \equiv \frac{1}{2} {\dot{\phi}}^2+\frac{1}{2}
 {\dot{\chi}}^2+\frac{\lambda}{4} \phi^4+\frac{g^2}{2} \phi^2 \chi^2. \label{model8}
\end{eqnarray}
Here $\rho_R$ is the energy density of the radiation.
We have included the radiation fluid in order to realize the radiation
dominated universe after preheating.
$\Gamma$ is a decay rate from scalar fields to the radiation.
We assume that both the decay rates of $\phi$ and $\chi$ to the radiation 
are the same for simplicity.

\section{Numerical results}
\subsection{Dependence of $\delta N$ on the initial value of $\chi$.}
In order to evaluate curvature perturbations generated during preheating
by using the $\delta N$ formalism,
we first study the dependence of the $e$-folding number $N$ until the Hubble 
parameter $H$ reaches a given value $H_{\rm fin}$ varying the initial values of fields,
that is,
fields values at the end of inflation. 
Perturbation of $\phi$ approximately corresponds to the adiabatic perturbation 
in the linear regime of preheating.
Hence the variation of $\phi$ at the end of inflation does not newly-generate 
curvature perturbations during preheating and we only vary the initial 
value of $\chi$ in the numerical simulations.

\begin{figure*}[t]
\begin{center}
  \includegraphics[width=16.cm,clip]{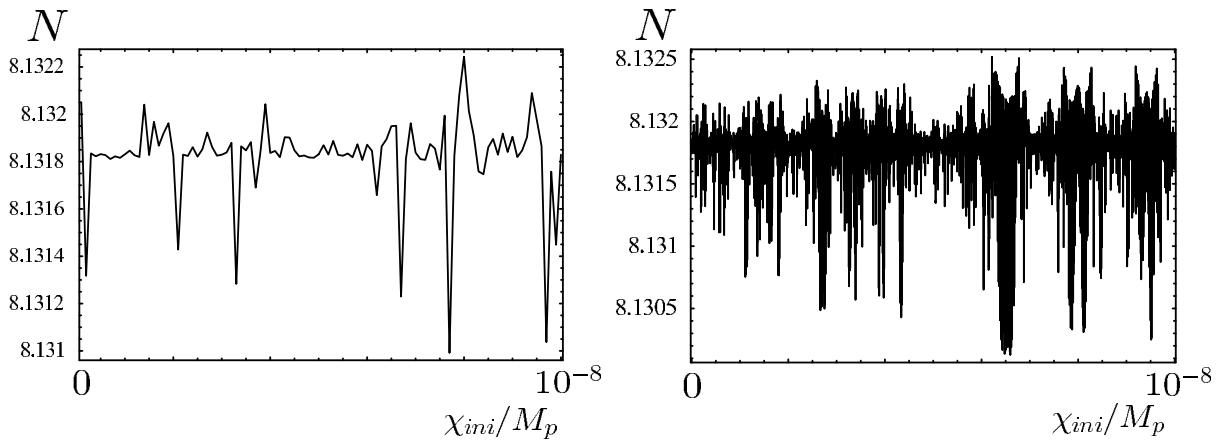}
\caption{Plots showing the initial value dependence of the $e$-folding number $N$.
In left and right panel,
100 and 10000 discrete points are plotted respectively within the same interval 
of the horizontal axis.
The parameters used in both panels are
$g^2 =2\lambda =0.04, \ \phi_{\rm ini}=M_p, {\dot \phi}_{\rm ini}={\dot \chi}_{\rm ini}=0, \Gamma=10^{-4}H_{\rm ini}, H_{\rm fin}=10^{-1}\Gamma$. 
}
\label{pa25a+26a}
\end{center} 
\end{figure*}

We show in Fig.~\ref{pa25a+26a} plots of $e$-folding number as a function of
the initial value of $\chi$ which we denote as $\chi_{\rm ini}$ for two different
values of the sample size of $\chi_{\rm ini}$ ($10^2$ and $10^4$ for the left and right 
panel respectively).
The parameters used in the simulations are
$g^2 =2\lambda =0.04, \ \phi_{\rm ini}=M_p, {\dot \phi}_{\rm ini}={\dot \chi}_{\rm ini}=0, \Gamma=10^{-4}H_{\rm ini}, H_{\rm fin}=10^{-1}\Gamma$. 

From both plots,
we find that the $e$-folding number $N$ is very sensitive to $\chi_{\rm ini}$.
This feature is also seen in \cite{Tanaka:2003ck} though the sample size is limited to
$10^2$ and the parameters such as decay rate $\Gamma$ are different from 
those used in our simulations.
Hence we expect that this feature is characteristic of preheating itself. 
We have increased the sample size to $10^6$ (we cannot show the plot in this case
because the file size becomes too large.) and found similar feature to the case of
the smaller sample size.

\begin{figure*}[t]
\begin{center}
  \includegraphics[width=16.cm,clip]{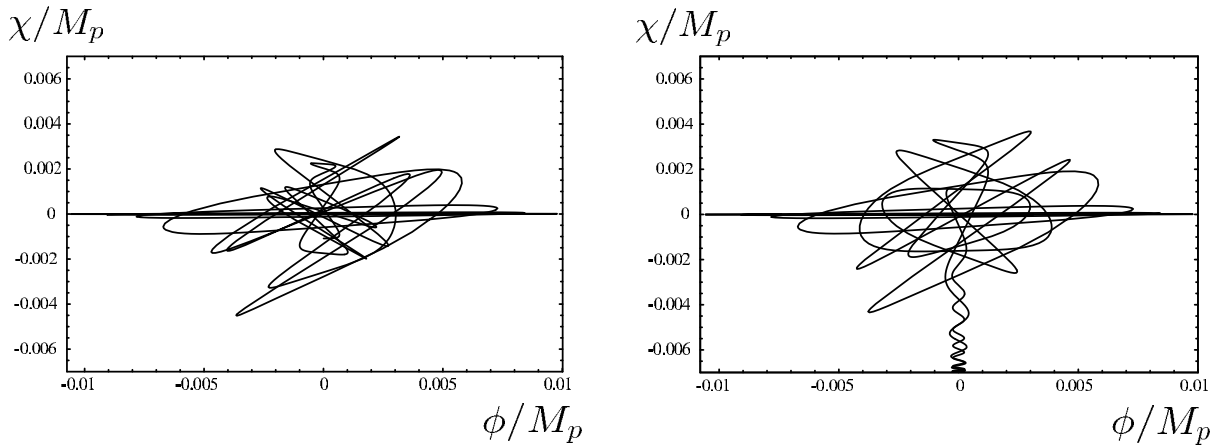}
\caption{Plots of the background trajectories of $\phi$ and $\chi$ during preheating.
In left and right panel,
$\chi_{ini}$ is $5.1\times 10^{-9}M_p$ and $5.2\times 10^{-9}M_p$ with other parameters
fixed to the same values.
We find that two trajectories soon deviate and become chaotic.
In left panel,
the background trajectory is trapped in the valley $\phi=0$ for a long time. 
}
\label{tra}
\end{center} 
\end{figure*}

This sensitive fluctuations of $N$ can be interpreted in a following way \cite{Tanaka:2003ck}.
Neglecting the radiation to which both $\phi$ and $\chi$ decay for simplicity,
the total energy density is given by $\rho_{\rm tot}=\rho_S$, and 
the evolution equation of $\rho_{\rm tot}$ is
\begin{equation}
\frac{d\rho_{\rm tot}}{dN}=-4\rho_{\rm tot}
-H\frac{d}{dN} \left( \phi_i {\dot \phi_i} \right)-3H \phi_i {\dot \phi_i}, \label{nu0.1}
\end{equation}
where $i$ labels $\phi$ and $\chi$.
In the oscillating phase after inflation,
the second and third terms oscillate.
If the period of these oscillations are much smaller than the Hubble time,
the averaged energy density over the period will decay in proportion to $e^{-4N}$,
like the radiation energy density.
In such a case, the evolution of the universe can be described 
by only one parameter, such as a radiation energy density, 
and the isocurvature perturbations do not arise. 
Hence additional curvature perturbations are not generated during preheating on 
super-horizon scales.
However, 
the sign of $\phi_i {\dot \phi_i}$ remains the same for a long time when the 
background trajectory is trapped in the potential valley along $\phi=0$.
We plot in Fig.~\ref{tra} the background trajectories of scalar fields for two 
slightly different initial values of $\chi$ (left panel is $\chi_{\rm ini}=5.1 \times 10^{-9}M_p$ 
and right one is $\chi_{\rm ini}=5.2 \times 10^{-9}M_p$) with other parameters fixed to
the same values.
We see that these two trajectories soon deviate and each trajectory becomes 
chaotic \cite{Jin:2004bf}.
In particular,
the behavior of $\phi_i {\dot \phi_i}$ is quite sensitive to the tiny difference of
$\chi_{\rm ini}$.
We found that when the background trajectory is trapped in the potential valley
along $\phi=0$,
which is seen remarkably in the right panel,
the time interval between two zeros of $\phi_i {\dot \phi_i}$ becomes so long that
it is smaller than the Hubble time only by a factor.
In such a case,
the expansion rate of the universe deviates significantly from that of the radiation 
dominated universe.
Because the background trajectory is chaotic,
the entrapment of the trajectory in the potential valley along $\phi=0$ occurs
almost randomly for $\chi_{\rm ini}$.
Hence the dependence of $N$ on $\chi_{\rm ini}$ becomes almost random.

\begin{figure*}[t]
\begin{center}
  \includegraphics[width=16.cm,clip]{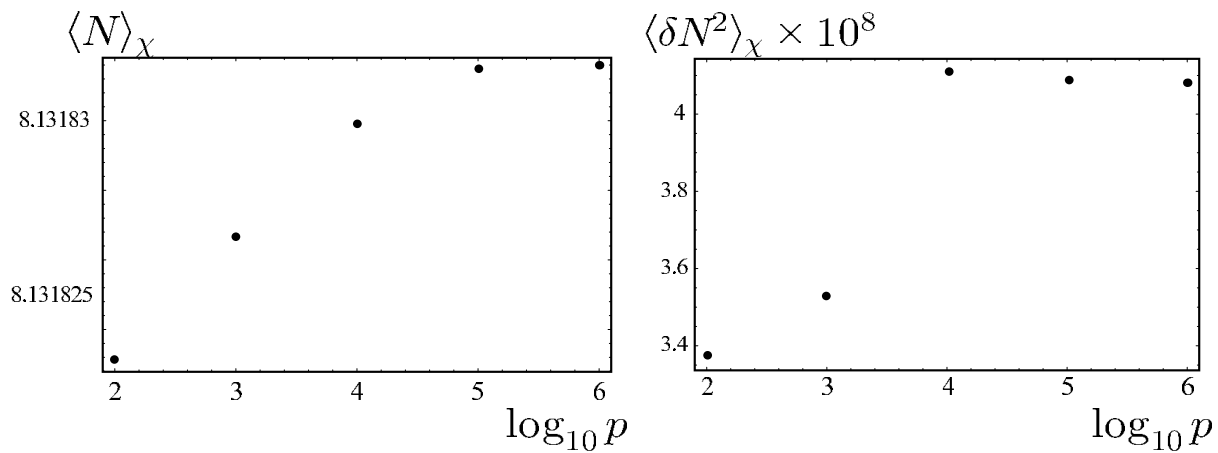}
\caption{Plots showing the dependence of the mean (left panel) and the 
variance (right panel) of $N$ on the sample size $p$.
We see that both the mean and the variance converge for large sample size.}
\label{mandv}
\end{center} 
\end{figure*}

Before we go to the discussion of how to obtain the power spectrum of curvature 
perturbations from the data which gives the correspondence between
$\chi_{\rm ini}$ and the $e$-folding number $N$,
we show in Fig.~\ref{mandv} the dependence of the mean and the variance of $N$ on 
the sample size, 
which we denote as $p$.
From this result,
we find that the dependence of both $\langle N(\chi) \rangle_{\chi} $ and 
$\langle {\delta N}(\chi)^2 \rangle_{\chi}$ on $p$ becomes weaker as $p$ becomes larger.
Here $\delta N (\chi) $ is defined by
\begin{equation}
\delta N (\chi) \equiv N (\chi)- \langle N(\chi) \rangle_{\chi}, \label{nu1}
\end{equation}
and $\langle \cdots \rangle_{\chi}$ is the average over the sampling points 
of $\chi_{\rm ini}$.
In particular,
the fluctuations of $\langle \delta N(\chi) \rangle_{\chi}$ for different values of $p$ are
within the statistical ones $\sim \sqrt{\langle \delta N(\chi)^2 \rangle_\chi /p}$,
which suggests that both the mean and the variance converge in proportion to $ 1/\sqrt{p}$
for $p \to \infty$ if we fix the range of $\chi$ in which we take average. 
Hence the smoothed $\delta N(\chi)$  over $\chi_{\rm ini}$ defined as
\begin{equation}
\delta N(\Delta \chi; \chi) \equiv \sum_{\chi_j} \delta \chi~ W_\chi (\Delta \chi; \chi_j-\chi) ~\delta N(\chi_j), \label{nu2}
\end{equation}
also converges if the large number of sampling points are contained in 
the range $(\chi-\Delta \chi, \chi+\Delta \chi)$. 
Here $\chi_j$ is the sampling point(integer $j$ labels the sampling point) 
and $\delta \chi \equiv \chi_{j+1}-\chi_j$.
$W_\chi (\Delta \chi; \chi)$ is the window function which rapidly drops for 
$|\chi| \gtrsim \Delta \chi$ and its normalization is defined by
\begin{equation}
\int d\chi ~W_\chi (\Delta \chi; \chi-\chi')=1. \label{nu2.1}
\end{equation}

In Fig.~\ref{pa25a+26a} we see that the standard deviation of $\delta N(\chi)$ 
which is about $2\times 10^{-4}$ is larger than the COBE normalization~\cite{Bunn:1996py}.
But this does not mean that curvature perturbations on large scales exceed
COBE normalization.
What we actually observe is the perturbation of $e$-folding number 
smoothed over many different spatial points in the real space,
or equivalently,
average of $\delta N(\chi)$ in $\chi_{\rm ini}$ space with some weight factor.
In the next subsection, we will show the relation between $\delta N(\chi)$ and 
${\cal P}_{\delta {\hat N}}(k)$.
Here we denote the perturbation of the $e$-folding number at a point $\vec{x}$ 
in the real space as $\delta {\hat N}(\vec{x})$ rather than $\delta N(\vec{x})$
in order to avoid confusion because we use $\delta N$ as the perturbation in $\chi$-field
space.

As we mentioned above,
$\delta N(\chi)$ is very sensitive to $\chi$ and the dependence of $\delta N(\chi)$ 
on $\chi$ seems almost random.
If this dependence is completely random,
then $\delta {\hat N}(\vec{x})$ is also random regardless of the shape of 
the power spectrum of $\chi$ at the end of inflation.
Hence, 
typical magnitude of $\delta {\hat N}(\vec{x})$ smoothed over the scale $R$ which is 
beyond the horizon scale at the time of preheating is given by the statistical 
fluctuations as
\begin{equation}
\sim \sqrt{ \langle {\delta N(\chi)}^2 \rangle_\chi } {\left( \frac{R}{r_H} \right)}^{-3/2}, \label{trend1}
\end{equation}
where $r_H$ is the horizon scale at the 
time of preheating.
Hence the smoothed $\delta {\hat N}(\vec{x})$ is significantly suppressed and 
the generation of curvature perturbations on super-horizon scales does not occur.
On the other hand, 
if $\delta N(\chi)$ is not completely randomized,
then the amplitude of the smoothed $\delta {\hat N}(\vec{x})$ would be larger 
than that of the statistical fluctuations,
in which case it may be possible that curvature perturbations can be larger or 
comparable to the COBE normalization.

\begin{figure}[h]
\begin{center}
  \includegraphics[width=8.cm,clip]{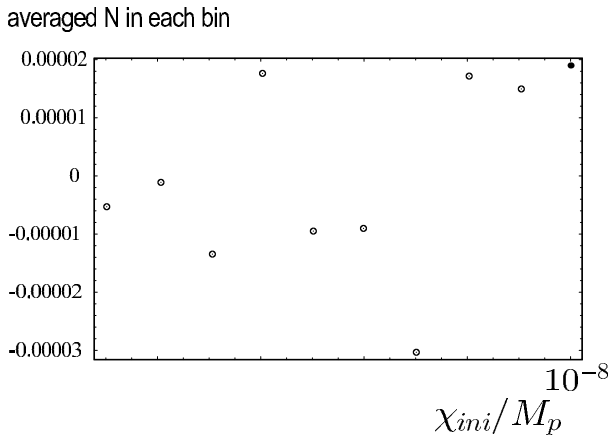}
\caption{Plot showing the $e$-folding number $N$ averaged in each bin
in which $10^5$ sampling points are contained.
We find that the fluctuation of the averaged $e$-folding number exceeds the
statistical fluctuation.
Hence $N$ is not completely randomized.
}
\label{ten-bin}
\end{center} 
\end{figure}

In order to see whether $\delta N(\chi)$ is randomized or not,
we divided the region of $\chi_{\rm ini}$ which contains one million sampling points 
into ten bins and took the average of $\delta N(\chi)$ for each bin with the same parameters
as in Fig.~\ref{pa25a+26a}.
Figure~\ref{ten-bin} is the result.
We see that the fluctuations exceed $10^{-5}$ by order of magnitude which is much larger 
than what is expected if $\delta N(\chi)$ is random.
Hence $\delta N(\chi)$ is not completely randomized.

\subsection{Construction of the Power Spectrum of $\delta {\hat N}$}

Generally, 
the power spectrum of the completely randomized field 
is proportional to $k^3$, that is, extremely blue \cite{Nambu:2005qh}.
In the previous subsection,
we have found that $\delta N(\chi)$ is not completely randomized,
so we expect that ${\cal P}_{\delta {\hat N}}(k)$ differs from the extremely 
blue power spectrum for sufficiently large scales if the isocurvature 
perturbations at the end of inflation,
which correspond to $\chi$-field perturbations in this model,
are not suppressed on large scales.
To see this,
we first define 
$\sigma (R)$ as the mean of the standard deviation of $\chi$ in the region of
size $R$,
\begin{equation}
\sigma^2 (R) \equiv \langle {\left( \chi (\vec{x})-\chi (R;\vec{x})\right) }^2 \rangle_R \approx \int_{R^{-1}}^{r^{-1}_H} \frac{dk}{k} {\cal P}_{\chi}(k), \label{ps9}
\end{equation}
where $\chi(R;\vec{x})$ is defined by
\begin{equation}
\chi(R;\vec{x}) \equiv \int d^3 x' ~W(R;\vec{x'}-\vec{x})\chi(\vec{x'})~.
\end{equation}
Here $W(R;\vec{x'}-\vec{x})$ is a window function which rapidly drops for
$|\vec{x'}-\vec{x}| \gtrsim R$ and its normalization is given by
\begin{equation}
\int d^3 x ~W(R;\vec{x'}-\vec{x})=1.
\end{equation}
We also define $\mu (R)$ as the standard deviation of $\chi$ smoothed over the scale $R$,
\begin{equation}
\mu^2 (R) \equiv \langle \chi^2 (R;\vec{x}) \rangle =\int_{L^{-1}}^{\infty} \frac{dk}{k} ~{| {\tilde W} (R^{-1};k) |}^2 {\cal P}_{\chi}(k),  \label{ps8} 
\end{equation}
where ${\tilde W}(R^{-1};k)$ is a Fourier window function and we have 
introduced a box of size $L(\gg R)$,
within which the stochastic properties are to be defined.

A key quantity for calculating the power spectrum of curvature perturbations 
on super-horizon scales is $\delta {\hat N}(\vec{x})$ smoothed over the scale $R >r_H$,
\begin{equation}
\delta {\hat N}(R;\vec{x}) = r^3_H \sum_{ \vec{x_i}} W(R; \vec{x_i}-\vec{x}) ( N(\vec{x_i})-\langle N \rangle_L), \label{rel1}
\end{equation}
where we have discretized the space into lattice whose grid spacing is $r_H$ and
have replaced the integral over the space with the sum over the lattice points 
$\vec{x_i}$ in the box.
$\langle \cdots \rangle_L$ denotes the expectation values inside the box.
In Eq.~(\ref{rel1}),
we implicitly assume that the spatial volume of size $R$ contains sufficient large 
number of lattice points so that the contamination of the random component is significantly 
reduced and $\delta N(R;\vec{x})$ does not depend on the size of $r_H$.
This assumption is valid as long as we consider sufficiently large scales
(but $R \ll L$ is imposed through).
 
In order to relate $\delta {\hat N}(R;\vec{x})$ to $\delta N(\chi)$ averaged 
over the sampling points of $\chi_{\rm ini}$,
we rewrite Eq.~(\ref{rel1}) as the sum over the sampling points on $\chi_{\rm ini}$ space,
\begin{equation}
\delta {\hat N}(R; \vec{x}) = {\left( \frac{r_H}{R} \right)}^3 \sum_{\chi_j} N (\chi_j) \left( n(R,\vec{x};\chi_j)-\langle n(R,\vec{x};\chi_j) \rangle_L \right), \label{nu2.4}
\end{equation}
where we have used the top hat window function.
$n(R, \vec{x};\chi_j)$ is a number of spatial points in the region
of size $R$ with $\vec{x}$ at its center in which $\chi_{\rm ini}$ takes 
the value between $\chi_j$ and $\chi_{j+1}$.
From the definition of $n(R, \vec{x};\chi_j)$,
we have the following relation,
\begin{equation}
\sum_{\chi_j} n(R,\vec{x};\chi_j)={\left( \frac{R}{r_H} \right)}^3. \label{nu3}
\end{equation}
From this relation, 
we can add the $\chi_j$-independent quantity to $N(\chi_j)$ in Eq.~(\ref{nu2.4}). 
So, 
as a matter of convenience,
we can rewrite Eq.~(\ref{nu2.4}) as
\begin{equation}
\delta {\hat N}(R; \vec{x}) = {\left( \frac{r_H}{R} \right)}^3 \sum_{\chi_j} \delta N (\chi_j) \left( n(R,\vec{x};\chi_j)-\langle n(R,\vec{x};\chi_j) \rangle_L \right). \label{nu2.5}
\end{equation}
Eq.~(\ref{nu2.5}),
or equivalently Eq.~(\ref{nu2.4}), 
is not strictly correct because from Eq.~(\ref{rel1}) to (\ref{nu2.5})
we have replaced $\chi (\vec{x_i})$ with $\chi_j$ which satisfies
$\chi_j <\chi (\vec{x_i}) < \chi_{j+1}$ and this will induce an error in
the corresponding replacement of $\delta N(\chi)$ of order $\sqrt{\langle \delta N^2 \rangle_\chi}$ 
for each lattice point. 
The sum of this error of $\delta N(\chi)$ over the lattice points would 
roughly correspond to the variation of $\langle \delta N(\chi) \rangle_{\chi}$ 
when each sampling point of $\chi_{\rm ini}$ is shifted by ${\cal O}(\delta \chi)$.
We have verified that the variation of $\langle \delta N(\chi) \rangle_{\chi}$ 
under the shift of $\chi_{\rm ini}$ by ${\cal O}(\delta \chi)$ lies within 
$\sqrt{ \langle \delta N(\chi)^2 \rangle_{\chi}}/\sqrt{p}$.
Hence Eq.~(\ref{nu2.5}) is correct to good accuracy as long as $R \gg r_H$. 

If $\delta \chi$ is small enough,
$n(R, \vec{x}; \chi)$ will be proportional to $\delta \chi$.
Hence we introduce $f(R,\vec{x};\chi)$ as
\begin{equation}
n(R,\vec{x};\chi)={\left( \frac{R}{r_H} \right)}^3 f(R,\vec{x};\chi) \delta \chi.
\end{equation}
Then $f(R,\vec{x};\chi)$ satisfies the following normalization condition,
\begin{equation}
\int d\chi~ f(R,\vec{x};\chi)=1. \label{rel4}
\end{equation}
Using Eq.~(\ref{nu2.5}),
we have
\begin{eqnarray}
\langle \delta {\hat N}^2(R;\vec{x}) \rangle_L &=& \sum_{\chi_j} \sum_{\chi_k} {(\delta \chi)}^2~ \delta N(\chi_j) \delta N(\chi_k) \nonumber \\
&&
\times \langle \big( f(R,\vec{x};\chi_j)-\langle f(R,\vec{x};\chi_j) \rangle_L \big) \nonumber \\
&&~
\times \big( f(R,\vec{x};\chi_k)-\langle f(R,\vec{x};\chi_k) \rangle_L \big) \rangle_L. \nonumber\\
\label{rel5}
\end{eqnarray}
We have to specify the function form of $f(R,\vec{x};\chi)$ which depends 
on the model in order to proceed further from Eq.~(\ref{rel5}).
We assume that $f(R,\vec{x};\chi)$ takes the form as,
\begin{equation}
f(R,\vec{x};\chi) = \frac{1}{ \sqrt{2\pi} \sigma (R)} \exp \left[ -\frac{1}{2 \sigma^2 (R)} {(\chi-\chi (R; \vec{x}) )}^2 \right]. \label{rel6}
\end{equation}
This assumption is valid if $\chi$-field obeys the Gaussian distribution,
which is the case if $\chi$-field is in the vacuum state at the end of inflation.
Under this assumption,
we have
\begin{eqnarray}
&&\langle f(R,\vec{x};\chi_j) \rangle_L = \frac{1}{ \sqrt{2\pi} \sigma (R)}\langle e^{- \frac{1}{2\sigma^2 (R)} {(\chi_j-\chi (R; \vec{x}))}^2 } \rangle_L, \label{rel7.5} \\
&&\langle f(R,\vec{x};\chi_j) f(R,\vec{x};\chi_k) \rangle_L = \frac{1}{2\pi \sigma^2 (R)} \langle e^{- \frac{1}{2\sigma^2 (R)} {(\chi_j-\chi (R; \vec{x}))}^2 } \nonumber \\
&& \hspace{40mm} \times e^{- \frac{1}{2\sigma^2 (R)} {(\chi_k-\chi (R; \vec{x}))}^2 } \rangle_L. \label{rel8}
\end{eqnarray}
We could replace ensemble average in Eqs.~(\ref{rel7.5}) and (\ref{rel8}) 
with the spatial average in the box of size $L$.
Because ${\vec x}$ appears in Eqs.~(\ref{rel7.5}) and (\ref{rel8}) only through
$\chi(R; \vec{x})$,
the ensemble average in these equations can be also interpreted as the ensemble 
average over $\chi (R; \vec{x})$ with some weight factor.
If $\chi$-field is the Gaussian random field,
then $\chi (R; \vec{x})$ is also Gaussian random field,
in which case the ensemble average of Eqs.~(\ref{rel7.5}) and (\ref{rel8}) 
becomes Gaussian integral over $\chi (R; \vec{x})$.
Then we have
\begin{eqnarray}
&&\langle e^{- \frac{1}{2\sigma^2 (R)} {(\chi_j-\chi (R; \vec{x}))}^2 } \rangle_L \nonumber \\
&&=\frac{1}{\sqrt{2\pi}\mu(R)} \int_{-\infty}^{\infty} d\chi (R; \vec{x}) e^{-\frac{\chi^2 (R,\vec{x})}{2\mu^2 (R)}} e^{- \frac{1}{2\sigma^2 (R)} {(\chi_j-\chi (R; \vec{x}))}^2 } \nonumber \\
&&=\frac{\sigma(R)}{\sqrt{\sigma^2(R)+\mu^2(R)}} e^{-\frac{\chi^2_j}{2(\sigma^2(R)+\mu^2(R))}}, \label{rel8.1} 
\end{eqnarray}
and
\begin{eqnarray}
&&\langle e^{- \frac{1}{2\sigma^2 (R)} {(\chi_j-\chi (R; \vec{x}))}^2 } e^{- \frac{1}{2\sigma^2 (R)} {(\chi_k-\chi (R; \vec{x}))}^2 } \rangle_L \nonumber \\
&&=\frac{1}{\sqrt{2\pi}\mu(R)} \int_{-\infty}^{\infty} d\chi (R; \vec{x}) ~e^{-\frac{\chi^2 (R; \vec{x})}{2\mu^2 (R)}}\nonumber \\
&& \hspace{10mm} \times e^{- \frac{1}{2\sigma^2 (R)} {(\chi_j-\chi (R; \vec{x}))}^2 } e^{- \frac{1}{2\sigma^2 (R)} {(\chi_k-\chi (R; \vec{x}))}^2 } \nonumber \\
&&=\frac{\sigma(R)}{\sqrt{\sigma^2(R)+2\mu^2(R)}} e^{-\frac{\mu^2(R)}{2\sigma^2(R) (\sigma^2(R)+2\mu^2(R))} {(\chi^2_j-\chi^2_k)}^2}\nonumber \\
&&\times e^{-\frac{\chi^2_j+\chi^2_k}{2(\sigma^2(R)+2\mu^2(R))}}. \label{rel8.2}
\end{eqnarray}
Substituting Eqs.~(\ref{rel8.1}) and (\ref{rel8.2}) into Eq.~(\ref{rel5}),
we have
\begin{eqnarray}
&&\langle \delta {\hat N}^2(R;\vec{x}) \rangle_L = \sum_{\chi_j} \sum_{\chi_k} {(\delta \chi)}^2 ~\delta N(\chi_j) \delta N(\chi_k) \nonumber \\
&&\times \Big[ \frac{1}{2\pi \sigma (R) \sqrt{ \sigma^2 (R)+2\mu^2 (R)}} e^{-\frac{\mu^2(R)}{2 \sigma^2 (R) (\sigma^2 (R) +2\mu^2 (R))} {(\chi_j-\chi_k)}^2} \nonumber \\
&&\times e^{-\frac{\chi_j^2+\chi_k^2 }{2 (\sigma^2 (R) +2\mu^2 (R))} }-\frac{1}{2\pi ( \mu^2 (R)+\sigma^2 (R) )} e^{-\frac{\chi_j^2+\chi_k^2}{2 (\sigma^2(R)+\mu^2(R))}} \Big]. \nonumber \\
\label{rel10}
\end{eqnarray}
Eq.~(\ref{rel10}) relates the power spectrum of curvature perturbations to 
the average of $ \delta N(\chi_j) \delta N(\chi_k)$ over $\chi_j$ and $\chi_k$ with some weight factor.

If the power spectrum of $\chi$-field at the end of inflation is red,
we can reduce Eq.~(\ref{rel10}) to a simpler form.
In this case,
we have $\mu (R) \gg \sigma (R)$.
Then Eq.~(\ref{rel10}) can be approximated as
\begin{eqnarray}
&& \langle \delta {\hat N}^2(R;\vec{x}) \rangle_L \approx \sum_{\chi_j} \sum_{\chi_k} {(\delta \chi)}^2 ~\delta N(\chi_j) \delta N(\chi_k) \nonumber \\
&&\times \Big[ \frac{1}{2 \sqrt{2} \pi \sigma (R) \mu (R)} e^{-\frac{1}{4 \sigma^2 (R)} {(\chi_j-\chi_k)}^2-\frac{1}{4 \mu^2 (R)} (\chi_j^2+\chi_k^2)} \nonumber \\
&&\hspace{10mm}-\frac{1}{2\pi \mu^2 (R)}e^{-\frac{1}{2\mu^2(R)} (\chi_j^2+\chi_k^2)} \Big] \nonumber \\
&&= \int_{-\infty}^{\infty} d\chi ~\frac{1}{ {(2\pi)}^{3/2} \sigma^2 (R) \mu (R)} \sum_{\chi_j} \sum_{\chi_k} \delta \chi^2 ~\delta N(\chi_j) \delta N(\chi_k) \nonumber \\
&& \hspace{10mm} \times e^{-\frac{1}{2 \sigma^2 (R)} {(\chi-\chi_j)}^2-\frac{1}{2 \sigma^2 (R)} {(\chi-\chi_k)}^2-\frac{1}{4 \mu^2 (R)} (\chi_j^2+\chi_k^2)} \nonumber  \\
&& \hspace{10mm} -{\Bigg[ \sum_{\chi_j} \delta \chi \frac{1}{\sqrt{2\pi}\mu (R)}e^{-\frac{1}{2\mu^2 (R)}\chi_j^2}\delta N(\chi_j) \Bigg]}^2 \nonumber \\
&& \approx \int_{-\infty}^{\infty} d\chi ~\delta N^2(\sigma (R);\chi) ~\frac{1}{\sqrt{2\pi} \mu (R)} \exp \Big[ -\frac{\chi^2}{2 \mu^2 (R)} \Big] \nonumber \\
&& \equiv G(\sigma(R), \mu (R)). \label{rel11}
\end{eqnarray}
Hence $\langle \delta {\hat N}^2(R;\vec{x}) \rangle_L$ is equal to the variance of 
$\delta N(\sigma(R);\chi)$ in the range $|\chi| \lesssim \mu (R)$.
We expect that ${\cal P}_{\delta {\hat N}}(k)$ shows deviation from the extremely 
blue power spectrum for scales $k \lesssim R^{-1}_c$ where $R_c$ is the critical scale
at which the fluctuation due to the randomness 
$\sim {\left( \frac{r_H}{R} \right)}^3 \langle \delta N(\chi)^2 \rangle_\chi$
goes down to the value given by Eq.~(\ref{rel11}).

On the other hand,
we have
\begin{equation}
\frac{d}{dR} \langle \delta {\hat N}^2 (R;\vec{x}) \rangle_L = \frac{d}{dR} \int^{R^{-1}} \frac{dk}{k} {\cal P}_{\delta {\hat N}}(k) =-\frac{1}{R} {\cal P}_{\delta {\hat N}}(R^{-1}). \label{tilt4}
\end{equation}
From this simple relation between $\langle \delta {\hat N}^2(R;\vec{x}) \rangle_L$ 
and the power spectrum of $\delta {\hat N}(\vec{x})$,
we can make a connection between ${\cal P}_{\delta {\hat N}}$ and
$\delta N^2 (\sigma (R); \chi)$ given by Eq.~(\ref{rel11}).

Furthermore, in order to have an analytic relation between ${\cal P}_{\delta {\hat N}}$ and
$\delta N^2 (\sigma (R); \chi)$, 
we assume that ${\cal P}_{\chi} (k)$ is given by
\begin{equation}
{\cal P}_{\chi}(k)={\cal P}_o {\left( \frac{k}{k_o} \right)}^{-n_{\chi}}, ~~~~n_{\chi}>0. \label{tilt1}
\end{equation}
Then $\mu (R)$ and $\sigma (R)$ become
\begin{equation}
\mu^2 (R) \approx \frac{1}{n_{\chi}} {\cal P}_o {(k_o L)}^{n_{\chi}}
, ~~~~~\sigma^2 (R) \approx \frac{1}{n_{\chi}} {\cal P}_o {(k_o R)}^{n_{\chi}}, \label{tilt2}
\end{equation}
where $L$ is the size of the box as mentioned before.
We see that $\mu (R)$ does not depend on $R$ and $\sigma (R)$ is an increasing function
with respect to $R$.
Using Eq.~(\ref{tilt2}),
we have
\begin{equation}
\frac{d}{dR} \langle \delta {\hat N}^2 (R;\vec{x}) \rangle =\frac{n_\chi}{2R} \sigma (R) \frac{\partial}{\partial \sigma (R)} G(\sigma(R),\mu(R)). \label{tilt3}
\end{equation}
From Eqs.~(\ref{tilt3}) and (\ref{tilt4}),
we obtain the relation between ${\cal P}_{\delta {\hat N}}$ and $G(\sigma(R),\mu(R))$,
\begin{equation}
{\cal P}_{\delta {\hat N}} (R^{-1}) = -\frac{n_{\chi}}{2} \sigma (R) \frac{\partial}{\partial \sigma (R)} G(\sigma(R),\mu(R)). \label{tilt5}
\end{equation}
Hence we can obtain the power spectrum of curvature perturbations at any scale
if we know the dependence of $\sigma(R)$ on $G(\sigma(R),\mu(R))$.

\begin{figure*}[t]
\begin{center}
  \includegraphics[width=17.cm,clip]{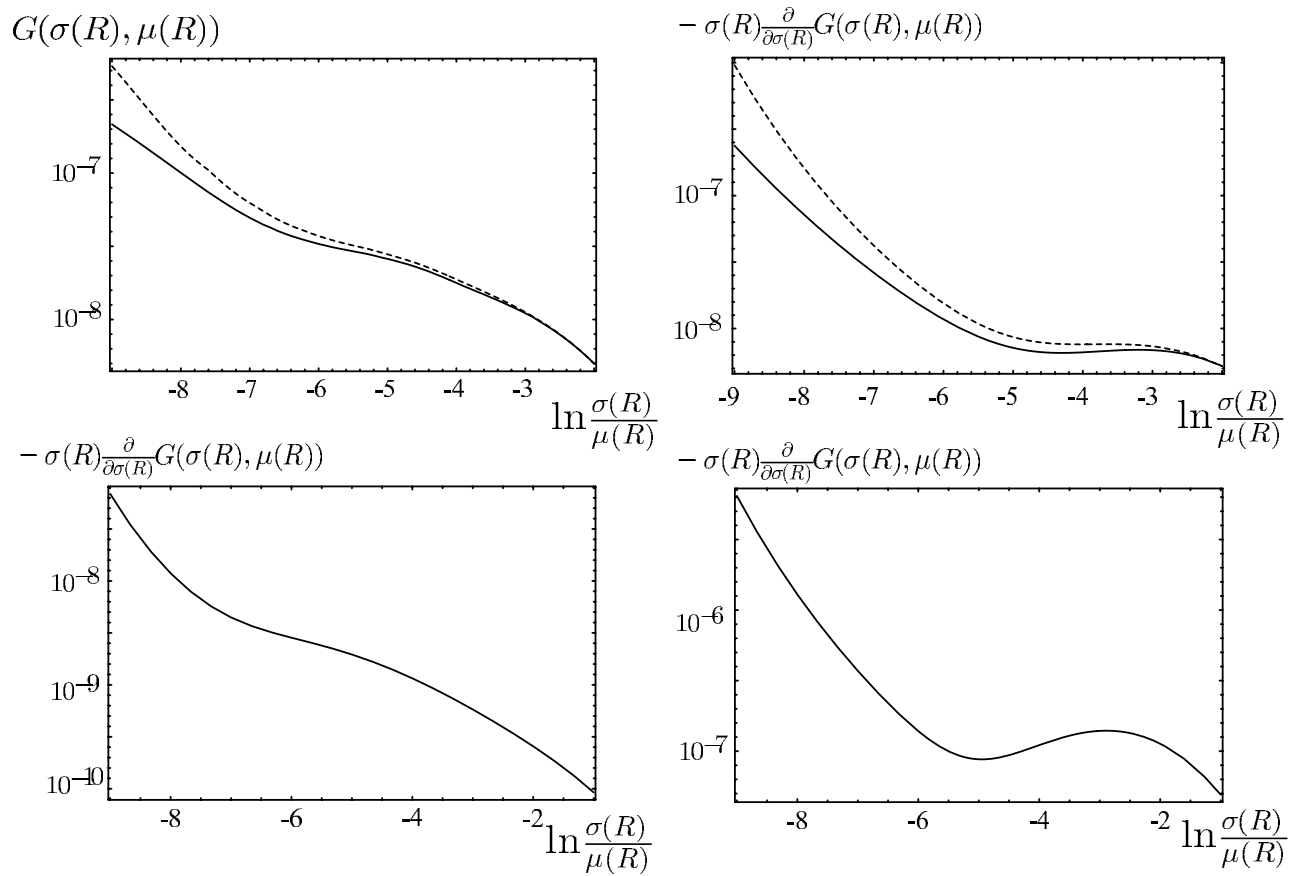}
\caption{In the top left panel,
$G(\sigma(R),\mu(R))$ is plotted as a function of $\sigma(R)$.
Each solid and dotted line corresponds to the case of $10^5$ and $10^6$
sampling points respectively.
The parameters are the same as used in Fig.~\ref{pa25a+26a}.
The remaining three panels plot 
$-\sigma(R) \frac{\partial}{\partial \sigma(R)} G(\sigma(R),\mu(R))$ as a function
of $\sigma (R)$ with $\mu(R)$ fixed,
hence showing the scale dependence of ${\cal P}_{\delta \hat{N}}$.
$\mu(R)$ in the top right, bottom left and bottom right is $5\times 10^{-5}M_p$,
$10^{-8}M_p$ and $10^{-2}M_p$,
respectively.
}
\label{sigma}
\end{center} 
\end{figure*}
 
The top left panel in Fig.~\ref{sigma} plots $G(\sigma(R),\mu(R))$ 
as a function of $\sigma(R)$ for fixed $\mu(R)$.
In this plot,
we have approximated $G(\sigma(R),\mu(R))$ as
\begin{equation}
G(\sigma(R),\mu(R)) \approx \frac{1}{2\mu(R)} \int_{-\mu(R)}^{\mu(R)} d\chi ~\delta N^2 (\sigma(R);\chi), \label{tilt6}
\end{equation}
for simplicity.
We assume that this replacement for calculating $G(\sigma(R),\mu(R))$ does not
cause significant error on the evaluation of the amplitude of curvature perturbations.
$\mu(R)$ is chosen to be $5\times 10^{-5} M_p$.
Solid line corresponds to the case of $10^6$ sampling points of $\chi_{\rm ini}$
within the region $0<\chi_{\rm ini}<\mu(R)$ and dotted one corresponds to the case of $10^5$
sampling points.

We first see that slopes of both two lines become steep at small $\sigma(R)$.
This steep rise is due to the contamination of the random component.
Since the sampling number in the width of $\mu(R)$ is limited in our 
simulations (at most $10^6$),
the sampling number in the width of $\sigma(R)$ becomes at most 
$\sim 10^6 \times \frac{\sigma(R)}{\mu(R)}$.
Hence if we decrease $\sigma(R)$ with $\mu(R)$ fixed,
the sampling number within the width of $\sigma(R)$ also decreases.
Below some critical $\sigma(R)$,
the sampling number is so small that  $\delta N(\sigma(R);\chi)$ is still 
dominated by the random component.  
In Fig.~\ref{sigma}, 
we actually observe that if we increase the number of sampling points
the amplitude of $G(\sigma(R),\mu(R))$ at small $\sigma(R)$, 
where the slope of the line is steep,
decreases and the difference between two lines becomes negligible for large $\sigma(R)$. 
Hence this steep rise at small $\sigma(R)$ is meaningless in a sense that we cannot 
obtain the correct amplitude of curvature perturbations at the corresponding scale $R$.
We have nothing left except to increase the sampling number in order to reduce the
contribution from the random component and obtain the correct $G(\sigma(R),\mu(R))$
at small $\sigma(R)$.
On the other hand,
at large $\sigma(R)$, 
we also see that the slopes become steep.
However,
this feature cannot be trusted because the approximation we made in Eq.~(\ref{rel11}) 
is no longer valid for $\sigma(R)/\mu(R)\sim {\cal O}(1)$.
In such a case,
we have to go back to Eq.~(\ref{rel10}),
which requires a heavy numerical calculation.

Thus, 
in the following discussion, 
when we discuss the property of the power spectrum,
we consider only for the scales corresponding to $\sigma(R)/\mu(R)\sim e^{-5}-e^{-3}$.

The remaining three panels(top right, bottom left and bottom right) plot 
$-\sigma(R) \frac{\partial}{\partial \sigma(R)} G(\sigma(R),\mu(R))$ which is 
proportional to ${\cal P}_{\delta \hat{N}}$ for three different values of $\mu(R)$ 
with other parameters such as $g^2/\lambda$ fixed to the same as in Fig.~\ref{pa25a+26a}.
$\mu(R)$ in the top right panel is the same as in the top left panel and 
the solid and the dotted lines in the right panel just correspond to the same type
of lines in the left panel.
$\mu(R)$ in the bottom left and right panels are $10^{-8}M_p$ and $10^{-2}M_p$ respectively.
We find that the amplitude and shape of the plot depend on $\mu(R)$.
In the top right panel,
$-\sigma(R) \frac{\partial}{\partial \sigma(R)} G(\sigma(R),\mu(R))$ remains almost constant
except for the region of small $\sigma(R)$ where the slope becomes steep due to the small
size of the sampling number and the region where $\sigma(R)/\mu(R) ={\cal O}(1)$.
On the other hand,
$-\sigma(R) \frac{\partial}{\partial \sigma(R)} G(\sigma(R),\mu(R))$ is a monotonically 
decreasing function in the bottom left panel and monotonically increasing function in
the bottom right panel.
 
Later,
we will give a qualitative explanation why these plots depend on $\mu(R)$.

Now let us consider what type of the power spectrum we can obtain from the numerical results above.
Since $\sigma(R)$ is an increasing function with respect to $R$,
going to larger scales means increasing $\sigma(R)$.
From Eq.~(\ref{tilt2}),
the horizontal axis of all the panels in Fig.~\ref{sigma} is
\begin{equation}
\ln \frac{\sigma(R)}{\mu(R)}=\frac{n_\chi}{2}\ln \frac{R}{L}. 
\end{equation}
Hence if we increase the scale of interest $R$ by a factor of 10,
then the corresponding point on the horizontal axis shifts by $\sim n_\chi$.
Since cosmologically interesting scales range from $10^4 {\rm Mpc}$ to $1 {\rm Mpc}$,
the corresponding range in the horizontal axis, 
which we denote as $\Delta \ln \frac{\sigma(R)}{\mu(R)}$, 
is $\sim 4n_\chi$.
From Eq.~(\ref{tilt5}),
the amplitude of ${\cal P}_{\delta {\hat N}}$ is determined by multiplying 
$-\sigma(R) \frac{\partial}{\partial \sigma(R)} G(\sigma(R),\mu(R))$
by $\frac{n_{\chi}}{2}$.
Therefore by choosing suitable $n_{\chi}$,
we can set ${\cal P}_{\delta {\hat N}}$ to ${\cal P}_{{\cal R}} \approx 2\times 10^{-9}$ 
corresponding to the COBE normalization.
For example,
if $\mu(R)=5\times 10^{-5}M_p$,
which corresponds to the case of the top right panel,
$-\sigma(R) \frac{\partial}{\partial \sigma(R)} G(\sigma(R),\mu(R))$
is about $10^{-8}$ in the range $-5 \lesssim \ln \sigma(R)/\mu(R) \lesssim -3$.
Hence if $n_\chi \approx 0.4$,
then ${\cal P}_{\delta {\hat N}}$ becomes $2\times 10^{-9}$, 
which corresponds to the COBE normalization.
In this case,
we have also $\Delta \ln \frac{\sigma(R)}{\mu(R)} \approx 1.6$.
If we assume that $L$ has a value so that the horizon scale $\sim 10^4 {\rm Mpc}$ 
corresponds to $\ln \frac{\sigma(R)}{\mu(R)}\sim -3$,
then $\Delta \ln \frac{\sigma(R)}{\mu(R)}$ lies in the range 
$-5 \lesssim \ln \sigma(R)/\mu(R) \lesssim -3$,
in which case ${\cal P}_{\delta {\hat N}}$ becomes scale invariant over
the cosmological scales.

In the same way as above,
for the bottom right panel,
if $n_\chi \approx 4\times 10^{-2}$,
then ${\cal P}_{\delta {\hat N}}$ gives the COBE normalization. 
In this case,
$\Delta \ln \frac{\sigma(R)}{\mu(R)} \approx 0.2$.
Because ${\cal P}_{\delta {\hat N}}$ changes little in this interval,  
the spectrum becomes almost scale invariant over the cosmological scales.
For the bottom left panel,
if we choose $n_\chi$ so that ${\cal P}_{\delta {\hat N}}$ gives the 
COBE normalization,
${\cal P}_{\delta {\hat N}}$ becomes too blue over the cosmological scales
and hence this case is excluded.

From Fig.~\ref{sigma},
we see that there is a tendency that the spectrum of the curvature perturbation 
becomes blue as $\mu(R)$ decreases.
This can be explained if we assume that $\delta N(\chi)$ becomes more random 
as $\mu(R)$ becomes smaller, because the power spectrum of the completely randomized field is 
extremely blue as mentioned before \cite{Nambu:2005qh}.
In our numerical calculation, 
we devided the range $(0,\mu(R))$ in $\chi_{\rm ini}$ space into ten bins and
took the average of $\delta N(\chi)$ for each bin (cf. see the Fig.~\ref{ten-bin}).
Then we verified that the deviation of fluctuation of averaged $\delta N(\chi)$ 
in each bin from the statistical fluctuation becomes smaller as $\mu(R)$ becomes 
smaller.
This result shows that the assumption above is valid.

It seems unnatural that both the amplitude and the spectral index of 
${\cal P}_{\delta {\hat N}}$ depend on the size of the box $L$ within which stochastic 
properties are defined.
Introducing the infrared cutoff $L$ is necessary,
because without such a cutoff the expectation value of $\chi^2$ diverges.
We have assumed throughout the paper that the mean value of $\chi$ vanishes in the box.
Let us now prepare a box whose size is much smaller than $L$ but is still larger
than the scales of interest inside the box of size $L$.
Then the average of $\chi$ inside the small box does not vanish in general. 
As shown in \cite{Lyth:2006gd, Boubekeur:2005fj}, 
statistical quantity such as power spectrum defined in the small box is not
the same as the one defined in the large box of size $L$.
It is only when one averages over the small boxes inside the larger box of size $L$
that the statistical quantities become the same as the one defined in the large box.
Hence $L$,
the size of the box within which the average of $\chi$ vanishes,
appears as a parameter which in this case controls the shape of 
${\cal P}_{\delta {\hat N}}$.

Testing non-Gaussianity of the primordial fluctuations has become one of 
the most powerful probe that distinguishes various inflationary and post-inflationary 
models \cite{Bartolo:2004if}.
Recently the level of non-Gaussianity generated during preheating was investigated
in \cite{Jokinen:2005by}.
The authors in \cite{Jokinen:2005by} consider the same preheating model as in this
paper.
However in \cite{Jokinen:2005by},
the primordial curvature perturbations are assumed to be generated during inflation
while second order fluctuations which are responsible for the large 
non-Gaussianity are amplified during preheating.
Though it is shown in \cite{Jokinen:2005by} that massless preheating model considered 
in this paper is already ruled out from observations due to large non-Gaussianity,
we have to reanalyze the problem because the mechanism of generating the primordial 
curvature perturbations in \cite{Jokinen:2005by} is completely different from that 
in our paper.
We are now investigating the level of non-Gaussianity generated during preheating
by using the $\delta N$ formalism.

Interestingly, 
the dynamics of preheating is completely determined by $g^2/\lambda$,
not by $\lambda$ and $g$ independently.
All numerical results shown above is obtained for any value of $\lambda$ if
we assume $g^2/\lambda=2$.
Since $\lambda$ determines the energy scale of inflation,
$\lambda$ fixes the amplitudes of the primordial gravitational waves produced
during inflation.
On the other hand,
as we have shown above,
if the isocurvature perturbations on super-horizon scales are amplified at 
the end of inflation with suitable negative spectral index and infrared cutoff,
the primordial curvature perturbations with almost scale invariant and 
$\sim 10^{-5}$ amplitude can be generated during preheating.
Hence we can lower the tensor to scalar ratio by lowering $\lambda$,
that is,
the energy scale of inflation. 
In this case,
$\lambda \phi^4$ inflation model,
which is ruled out from the WMAP three year results 
combined with SDSS \cite{Spergel:2006hy, Tegmark:2001jh, Alabidi:2006qa, Kinney:2006qm, Martin:2006rs},
comes into the observationally allowed region.
\footnote{Although, see also \cite{Easther:2006tv}.}
Lowering the tensor to scalar ratio in the chaotic inflationary scenario was
also considered in \cite{Komatsu:1999mt}(see also \cite{Tsujikawa:2004my}),
though the mechanism is completely different from the one in this paper.

For completeness,
we consider the case in which the power spectrum of $\chi$ at the end of inflation
is blue,
that is,
$\mu(R) \ll \sigma (R)$.
In this case,
Eq.~(\ref{rel10}) can be approximated as
\begin{eqnarray}
&&\langle \delta N^2(R;x) \rangle \approx \sum_{\chi_j} \sum_{\chi_k} \delta \chi^2 \delta N(\chi_j) \delta N(\chi_k) \nonumber \\
&&\hspace{10mm} \times \frac{1}{2\pi \sigma^2 (R)} e^{-\frac{1}{2\sigma^2(R)} (\chi_j^2+\chi_k^2)} \Big[ e^{-\frac{\mu^2(R)}{2\sigma^4(R)} {(\chi_j-\chi_k)}^2}-1 \Big]. \nonumber \\
\label{rel12}
\end{eqnarray}
Because $\frac{\mu^2(R)}{\sigma^4(R)} {(\chi_j-\chi_k)}^2$ is much 
smaller than unity,
right hand side of Eq.~(\ref{rel12}) is suppressed.
Hence the generation of curvature perturbations on large scales does not
occur in this case.
In the scale-invariant case, 
that is, 
$\mu(R) \simeq \sigma(R)$,
we have no appropriate approximation in order to reduce Eq.~(\ref{rel10}) to 
a simpler form.
So, 
in this case, 
we have to calculate Eq.~(\ref{rel10}) directly.

\section{Conclusion}
The main purpose of this paper is to examine whether the generation of the 
curvature perturbations occurs or not during preheating and, 
if it occurs,
whether they could be seeds for large-scale structure in the universe. 
 
To do so, 
we applied $\delta N$ formalism to two-field preheating model whose potential 
is given by $\frac{\lambda}{4}\phi^4+\frac{g^2}{2}\phi^2 \chi^2$.
Then we have analyzed the dependence of the $e$-folding number during preheating 
on the isocurvature perturbations at the end of inflation which corresponds to 
perturbations of $\chi$-field.
We verified that the $e$-folding number during preheating is very sensitive to the 
initial value of $\chi$ due to the chaotic nature of the background trajectory
in field space,
which has been already observed in \cite{Tanaka:2003ck}.
Though the $e$-folding number seems almost random as a function of $\chi_{ini}$,
we found that the fluctuation of the smoothed $e$-folding number over a large 
number of initial values of $\chi$ exceeds the statistical fluctuation.
Hence the $e$-folding number is not completely randomized.

This deviation of the smoothed $e$-folding number from the statistical fluctuation
provides the generation of curvature perturbations on large scales.
We numerically found that the amplitude and the spectral index of the 
curvature perturbations originating from preheating at current cosmological 
scales can be fixed to the observed values by tuning the spectral index of 
$\chi$-field and the infrared cutoff.
An interesting point is that this consequence is derived only from the requirement
$g^2/\lambda=2$.
Hence by lowering $\lambda$ which determines the energy scale of inflation,
we can lower the tensor to scalar ratio to the observationally allowed value.

\acknowledgements

We would like to thank Takahiro Tanaka for continuously encouraging us to pursue 
our study and giving us useful comments.
We would like to thank also Keisuke Izumi, Takashi Nakamura, Misao Sasaki, Jiro Soda, 
Shinji Tsujikawa, Masahiro Yamaguchi and David H.~Lyth for useful comments.
The numerical calculations were carried out on Altix3700 BX2 at YITP in Kyoto University.

\end{document}